# Packaging of Low-Environmental-Sensitivity WGM Resonators for Practical Applications

Jiajun Wu, Xuanqi Wang, Chengyu Zhang, Chenghong Li, Shan Zhong, and Songbai Kang

*Abstract*—We presents a novel prism-coupled packaging strategy for WGM Resonators. Utilizing an all-solid-state optical adhesive process combined with active temperature control and hermetic sealing, the proposed package exhibits exceptional long-term stability and environmental robustness. For the first time, a standalone WGMR module was characterized, demonstrating a temperature sensitivity below $10^{-7}$ /°C and a low-frequency Z-axis acceleration sensitivity below $10^{-10}$ /g. Furthermore, the application of this module was explored as a stable optical frequency reference and a nonlinear photonic platform, achieving a short-term frequency stability of $2 \times 10^{-13}$ at 2 ms and generating Kerr soliton microcombs with a pump power of 100 mW. This compact, robust, and stable packaging solution significantly enhances the immediate applicability of WGMRs in real-world applications such as narrow-linewidth lasers and portable microcombs, thereby facilitating the transition of WGMR technology from laboratory research to practical deployment.

*Index Terms*—Whispering gallery mode resonator, packaging, temperature sensitivity, acceleration sensitivity, frequency reference, optical frequency combs

## I. INTRODUCTION

Whispering Gallery Mode Resonators (WGMRs), known for their compact size, high quality factor (Q), and small mode volume, are widely used in advanced laboratory applications such as optical sensing, narrow-linewidth lasers, frequency combs, and nonlinear optics [1-3]. However, WGMRs are highly sensitive to the environment: surface contaminants like moisture and dust can severely degrade their Q-factor, while temperature fluctuations and vibrations induce resonance frequency shifts. Moreover, laboratory work with WGMRs typically requires precise mechanical coupling fixtures, microscope alignment systems, and specialized operational skills. These constraints significantly limit their practical application beyond laboratory settings. Consequently, the development of a plug-and-play WGMR packaging technology — capable of ensuring stable coupling, offering a compact form factor, and maintaining a high Q-factor over the long term—is a crucial prerequisite for transitioning WGMRs from lab prototypes to real-world applications.

Current reported WGMR packaging predominantly employs tapered fiber coupling [6, 7]. However, this approach presents several practical limitations. First, It exhibits high sensitivity to the vibration-induced coupling fluctuations; second, The tapering process compromises the optical fiber's polarization-maintaining properties, degrading input light polarization purity and undermining long-term coupling stability[8]; last, tapered fiber coupling lacks capability for selective single-mode coupling adjustment[9]. The prism-coupling can resolve above limitations, but the implementation challenges persist: critical optical alignment for coupling into the resonator 's ultra small mode area and ultra-low aging for the displacements between the prism , WGMR, and collimators during adhesive curing[10]. While OEwaves Inc. has developed a commercial prism-coupled WGMR module [11], detailed technical specifications of its packaging scheme, comprehensive experimental characterization data, and in-depth exploration of its application potential have rarely been documented in the open scientific literature.

This paper presents a completely new WGMR packaging solution based on prism coupling. The assembly is secured using an all-solid-state optical adhesive bonding process, integrated with active temperature control and hermetic packaging, thereby achieving long-term coupling stability and exceptional environmental adaptability. Key performance parameters, including the Q-factor, coupling efficiency, insertion loss, and polarization extinction ratio, exhibited no significant degradation over a testing period exceeding six months. Furthermore, we conducted the first detailed investigation into the temperature and vibration characteristics of this packaged WGMR module based on Pound-Drever-Hall (PDH) technique. The application potential of the module was also demonstrated in two scenarios: as a stable optical frequency reference and as a platform for nonlinear Kerr soliton microcomb generation. These results indicate that the proposed packaging approach offers high stability, a compact form factor, cost-effectiveness, and user-friendliness, thereby facilitating the deployment of WGMRs in demanding field environments outside laboratory settings.

(Corresponding author: Songbai Kang.)

Jiajun Wu, Chengyu Zhang, Chenghong Li are with the Key Laboratory of Atomic Frequency Standards, Innovation Academy for Precision Measurement Science and Technology, Chinese Academy of Sciences, Wuhan 430071, China, and with the University of Chinese Academy of Sciences, Beijing 100049, China (e-mail:wujiajun19@mails.ucas.ac.cn; zhangchengyu24@mails.ucas.ac.cn; lichenhong19@mails.ucas.ac.cn).

Xuanqi Wang is with the Hubei Key Laboratory of Optical information and Pattern Recognition, Wuhan Institute of Technology, Wuhan 430205, China. (e-mail: a370884786@163.com).

Shan Zhong, Songbai Kang are with the Key Laboratory of Atomic Frequency Standards, Innovation Academy for Precision Measurement Science and Technology, Chinese Academy of Sciences, Wuhan 430071, China (e-mail: kangsongbai@apm.ac.cn; zhongshan@apm.ac.cn).

## II. PACKAGING AND CHARACTERISTICS.

A high-Q magnesium fluoride ($MgF_2$) WGMR was fabricated via single-point diamond turning. The resonator features a diameter of 6 mm and thickness of 0.5 mm (Fig. 1(a)). To suppress higher-order mode excitation, the rim width was optimized to about 25 μm, defining the effective coupling region. Simulation results (Fig. 1(a)) confirm strong optical confinement, showing a fundamental mode volume of about 40 μm². The critically coupled transmission spectrum of the bare resonator (Fig. 1(b)) exhibits a linewidth-derived Q-factor of about $2\times10^9$ at 1.5 μm.

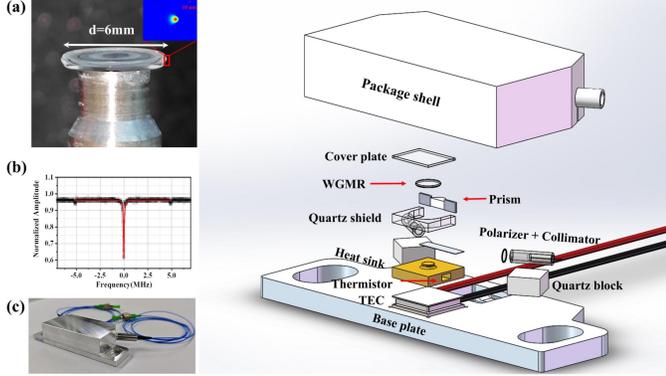

Fig. 1. This is a sample of a figure caption.Fig.1. (a) $MgF_2$ WGMR image and mode profile. (b)Cavity linewidth measurement at critical coupling. (c)Photo of WGMR packaging based on prism coupling. (d)Schematic diagram of the $MgF_2$ WGMR-prism coupling packaged structure.

The complete assembly (Fig. 1c) occupies approximately 73 cm³ while maintaining all critical optical functionalities. Fig. 1d illustrates the WGMR-prism coupling architecture. A precisely machined aperture in the metal base substrate houses a thermoelectric cooler (TEC) coupled to a tungsten-copper heatsink. The WGMR is mounted on the heatsink surface and hermetically sealed by the coupling prism, glass frame, and cover plate–effectively isolating the resonator from atmospheric particulates and humidity. Polarization-maintaining (PM) fiber collimators enable laser input/output coupling. A polarizer preceding the input collimator ensures >20 dB polarization extinction ratio for selective TE-mode excitation. Key optical components (WGMR, prism, collimators) are bonded with ultra-low thermal expansion UV adhesive using sequential UV curing followed by a final high-temperature cure to ensure coupling stability and residual stress dissipation. The integrated TEC can be used actively stabilizes resonator temperature.

Fig.2 (a) show that the coupling efficiency of four fabricated module samples remained stable over six months of operation, showing no degradation. Notably, the prism-coupled WGMR module achieved a coupling efficiency exceeding 90% (Fig.2 (b)) with an overall insertion loss of approximately 2 dB, which almost matches the performance of tapered fiber coupling approach. Fig.2 (c) presents the optical Q-factor of Module #4, measured using the ring-down technique after a six-month testing period. The experimental decay curve was fitted to obtain the final value.

Broad-wavelength laser scanning (about 0.5 nm range) performed on the WGMR module revealed the cavity resonance spectrum shown in Fig. 2(d). The observed spectrum presents that the WGMR has about 11. 5 GHz free spectrum range (FSR) and six modes per FSR (green annotation), indicating it is a few-mode WGMR. This demonstrates that effective mode suppression has been achieved through optimized cavity chamfer design and coupling angle adjustment. This clean mode feature is very attractive for ultra-high sensitivity sensing and particle detection, optical frequency reference, nonlinear optics and frequency conversion applications.

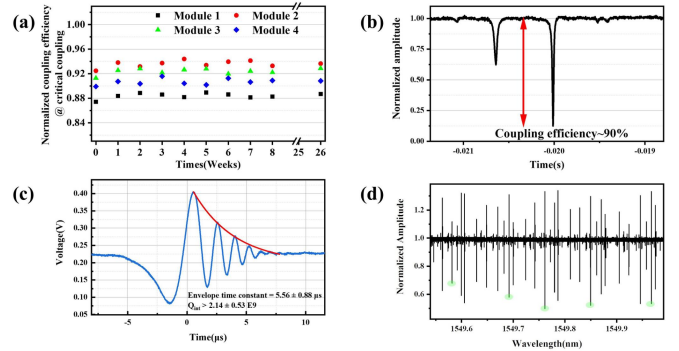

Fig. 2. (a) Long-term consistency test of the linewidth (Q-factor) for four modules. (b) Modules achieved a coupling efficiency of approximately 90%. (c) Intrinsic Q-factor measurement of Module #4 under critical coupling conditions. (d) Broad-range laser transmission spectrum of the WGMR module.

## III. TEMPERATURE SENSITIVITY.

Compared to low-expansion Fabry-Pérot resonators, crystalline WGMR exhibit significantly higher thermal sensitivity (about $10^{-5}$ / °C ) [12] due to pronounced thermal expansion effects. To suppress thermally induced frequency drift, we achieved active temperature control of the WGMR by designing and integrating a thermoelectric cooler (TEC) within the packaging module.

The temperature sensitivity testing system for the WGMR packaging module is shown in Fig. 3(a). The WGMR device was mounted on a heating plate, and subjected to stepwise temperature increments to simulate the environmental temperature variations. A commercial laser (NKT e15) was actively locked to the resonator 's one resonance via PDH stabilization. A high-resolution wavemeter (Bristol 621A) was used to monitor the laser frequency drift. Fig. 3 (b), (c) present the time-domain recordings of the temperature of the heating plate and WGMR 's temperature and laser output frequency without and with the active temperature stabilization. We can see that the WGMR temperature can be well controlled with the help of the integrated TEC controller when the temperature of the heating plate was regulated step by step. Furthermore, the simultaneously traced frequency data indicate that the bare $MgF_2$ WGMR' s temperature sensitivity is about -500 MHz/°C and the device 's temperature sensitivity is about +50 MHz/°C, representing a 10-fold reduction in environmental temperature sensitivity. We note the positive polarity of the temperature sensitivity of the device stems from the temperature gradient between the sensor location and the WGMR resonator. The rising ambient temperature reduced TEC heating power (at a fixed setpoint), thereby lowering the cavity temperature. Future design iterations focused on reducing thermal gradients are expected to significantly improve the cavity's temperature stability.

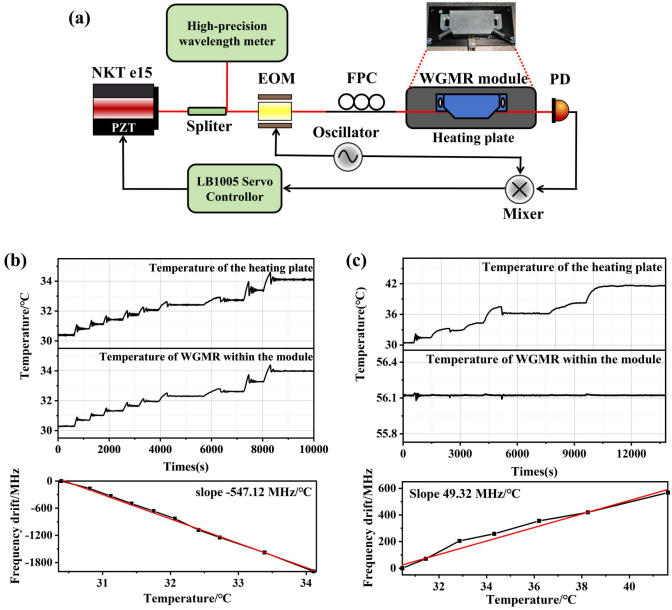

Fig. 3. Crystalline WGMR module cavity frequency temperature sensitivity characterization (a) Experimental setup; (b)Temperature sensitivity measurement of WGMR package during free-running; (c)Temperature sensitivity measurement of WGMR package under active temperature control.

To further investigate the device's long-term temperature feature, we measured the stabilized laser's frequency under the normal laboratory conditions without any auxiliary thermal management. Figs. 4(a) and (b) present the resonant frequency drift and long-term stability of the WGMR module, measured by the beat note between the locked laser and an H-maser-referenced optical frequency comb with/without active temperature stabilization. With active control, cavity frequency fluctuations remained ≤5 MHz over $10^4$ s ;Without control, the cavity drifted > 55 MHz within 2500 s, and easy to lose lock. Temperature stabilization enhanced frequency stability by over one order of magnitude beyond 100 seconds. The results confirmed that the packaging scheme significantly reduced the temperature sensitivity of WGMR and improved long-term frequency stability.

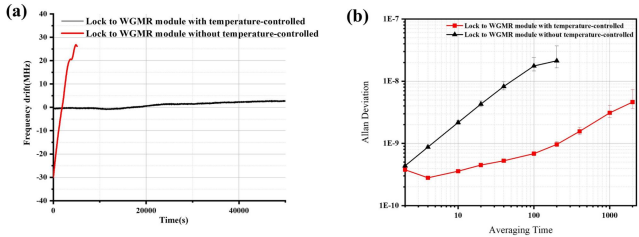

Fig. 4. (a) compares the frequency drift of the laser locked to the cavity with (black curve) and without (red curve) the internal With with active temperature control. (b) presents the corresponding frequency stability of the locked laser.

## IV. ACCELERATION SENSITIVITY.

Acceleration sensitivity is a key parameter for assessing the impact of environmental vibration on WGMR. However, experimental data to validate the theoretically predicted low sensitivity [13] remains scarce. Existing measurements are limited: the results from Wei Zhang et al. [14] were constrained by coupling instability, while the work by Anatoly Savchenkov et al. [15] reflected the combined response of the laser and the microresonator, not the WGMR's intrinsic acceleration sensitivity.

To quantify our WGMR packaging's acceleration sensitivity, a dedicated test setup was constructed (Fig.5(a)): Only the test module and an accelerometer were mounted on the vibration shaker, while all other components remained vibration-isolated. The laser (NKT e15) was locked to the WGMR with about 1 kHz bandwidth. Vibrations along the device's most sensitive axis (Z-axis) were applied from 10 to 100 Hz in 10 Hz increments, with synchronous recordings of (i) piezoelectric accelerometer data and (ii) phase noise of the locked laser.

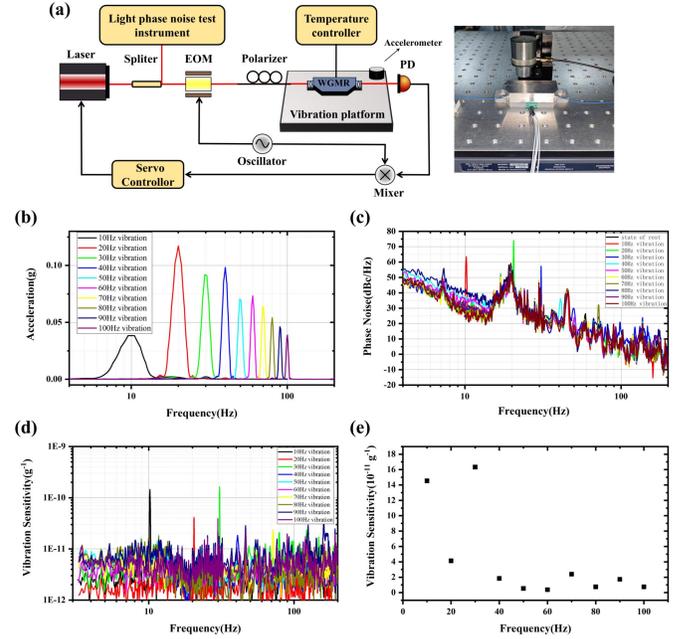

Fig. 5. Crystalline WGMR module cavity frequency acceleration sensitivity characterization (a) Experimental setup; (b)The acceleration applied to the WGMR module measured by the accelerometer; (c) Under vibration excitation, the phase noise of the WGMR-locked laser, measured by a phase noise characterization system; (d) Quantifies the acceleration sensitivity of the WGMR module, derived from vibration-induced laser phase noise conversion. (e)Extract the single-point acceleration sensitivity of the WGMR module From Figure (d).

Fig.5(b) displays the measured acceleration magnitude at each frequency. Fig.5(c) shows the phase noise spectra under corresponding vibrations, revealing that resonance frequency modulation induces characteristic peaks at offset frequencies. Following [15], the Z-axis acceleration sensitivity was derived from this data (plotted in Fig.5(c)). Notably, peak sensitivity reaches $10^{-10}$/g at 10 Hz and 30 Hz (likely due to package mechanical resonances), while maintaining <$10^{-11}$ /g at other frequencies. As extracted in Fig.5(d), acceleration sensitivity remains below $1.63 \times 10^{-10}$/g across 10 to 100 Hz, consistent with the theoretical prediction of about $10^{-10}$/g in Ref. [15].

## IV. Applications.

We demonstrate the dual functionality of a packaged WGMR module as a frequency reference and a nonlinear photonics platform. Using the PDH technique, we lock a homemade laser to the WGMR. The phase noise is suppressed by over 25 dB near 1 kHz offset, approaching the thermorefractive noise limit (Fig. 6a). The fractional frequency stability reaches $2 \times 10^{-13}$ at 2 ms integration time (Fig. 6b). Furthermore, with 100 mW input power, single-soliton generation is achieved (Fig. 6c, d). These results confirm the module's capability as a compact, low-cost, and robust platform for precision metrology and frequency comb generation.

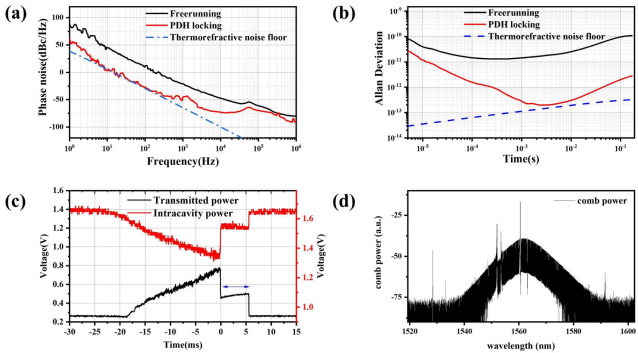

Figure 6 demonstrates the dual functionality of the WGMR module as an optical frequency reference and a nonlinear Kerr frequency comb source. (a) compares the phase noise of a domestic DFB laser in free-running operation (black curve) and when locked to the WGMR module via the PDH technique (red curve). (b) displays the corresponding frequency stability (Allan deviation) derived from the phase noise. The locked stability (red curve) reaches 2×10-13 at 2 ms. (c) and (d) show the optical spectra of the single-soliton state and Kerr soliton comb generated at an input power of 100 mW.

## V. Conclusion

We demonstrate, for the first time, a robust packaging technique for a prism-coupled $MgF_2$ WGMR, utilizing low-expansion optical adhesive and active temperature control in a sealed design. To accurately quantify the environmental sensitivity, a dedicated testbed was established. Precise measurements via the PDH technique revealed a thermal sensitivity below $10^{-7}$ /°C under active temperature control—a full order-of-magnitude improvement over the bare resonator—along with a Z-axis vibration sensitivity below $10^{-10}$ / g. The packaged module maintained stable coupling for six months and demonstrated dual functionality: enabling laser frequency stabilization at $2 \times 10^{-13}$ at 2 ms and generating Kerr solitons with only 100 mW input. This packaging approach effectively enhances immunity to temperature fluctuations and mechanical vibrations, improves operational reliability and lifetime, and facilitates the transition of WGMR technology from laboratory research to large-scale practical applications.